\documentclass[epj]{svjour}

\usepackage{graphicx}
\usepackage{fancyhdr}
\usepackage{amssymb}
\usepackage{subfigure}
\usepackage{psfrag}
\usepackage{axodraw}
\usepackage{amsmath}

\def\makeheadbox{{
 }}
\def\mytitle{$B_s-\bar{B}_s$ mixing and $B_s \to KK$ decays within supersymmetry}
\def\myauthors{Seungwon Baek}
\def\mytype{Contributed Talk}
\def\mysession{Flavor}

\newcommand{\be}{\begin{equation}}
\newcommand{\ee}{\end{equation}}
\newcommand{\bea}{\begin{eqnarray}}
\newcommand{\eea}{\end{eqnarray}}
\newcommand{\beq}{\begin{equation}}
\newcommand{\eeq}{\end{equation}}
\newcommand{\ba}{\begin{array}}
\newcommand{\ea}{\end{array}}
\newcommand{\bi}{\begin{itemize}}
\newcommand{\ei}{\end{itemize}}
\newcommand{\bn}{\begin{enumerate}}
\newcommand{\en}{\end{enumerate}}
\newcommand{\bc}{\begin{center}}
\newcommand{\ec}{\end{center}}
\newcommand{\de}{\delta}
\newcommand{\te}{\theta}
\renewcommand{\l}{\left}
\renewcommand{\r}{\right}

\newcommand{\ol}{\overline}

\newcommand{\nl}{\nonumber\\}

\newcommand{\wt}[1]{\widetilde{#1}}

\def\sss{\scriptscriptstyle}

\def\bd{B_d^0}
\def\bs{B_s^0}
\def\btopik{B \to \pi K}
\def\pewcp{P_{\sss EW}^{\prime\sss C}}
\def\pewp{P'_{\sss EW}}

\def\ApNPcomb{{\cal A}^{\prime, comb} e^{i \Phi'}}

\def\bs{B_s^0}
\def\kbar{{\bar K}^0}
\def\bskkneut{{\bs}\to K^0 \kbar}
\def\bskk{{\bs}\to K^+ K^-}

\begin{document}
\title{$B_s-\bar{B}_s$ mixing and $B_s \to KK$ decays within supersymmetry}
\author{Seungwon Baek\inst{1}
\thanks{\emph{Email:} sbaek@korea.ac.kr}%
}                     
%
%
\institute{The Institute of Basic Science and Department of Physics,
Korea University, Seoul 136-701, Korea
}
%
\date{}
\abstract{
We consider the constraint of $B_s-\bar{B}_s$ mass difference, $\Delta m_s$,
on an MSSM scenario with large flavor mixing. Even with this constraint,
we show that a large deviation from the SM in CP asymmetries of
$B_s \to KK$ decays is still possible, making this channel promising
in search for supersymmetry.
} 
\maketitle
%

\section{Introduction}
\label{intro}

The flavor changing processes in the ${s}-{b}$ sector are sensitive probe
of new physics (NP) beyond the standard model (SM) because they are experimentally 
the
least constrained.
In the minimal supersymmetric standard model (MSSM), however, the
flavor mixing in the chirality flipping
down-type squarks, $\wt{s}_{L(R)}-\wt{b}_{R(L)}$, is already strongly
constrained by the measurement of $BR(B \to X_s +\gamma)$. On the other hand,
large flavor mixing in the chirality conserving
$\wt{s}_{L(R)} - \wt{b}_{L(R)}$ has been largely allowed.
Especially the large mixing scenario in the $\wt{s}_R - \wt{b}_R$ sector
has been drawing much interest
because it is well motivated by the measurement large neutrino mixing
and the idea of grand unification~\cite{Baek:GUT}.

The D{\O} and CDF collaborations at Fermilab Tevatron reported the results
on the measurements of $B_s - \ol{B}_s$ mass difference~\cite{D0,CDF}
\bea
17 ~{\rm ps}^{-1} < \Delta m_s < 21 ~{\rm ps}^{-1} ~~(90 \% \mbox{~CL}), \nl
\Delta m_s = 17.33^{+0.42}_{-0.21}\pm 0.07 ~{\rm ps}^{-1},
\label{dms:exp}
\eea
respectively.
These measured values are consistent with the SM predictions~\cite{Bona:2005eu,CKMfitter}
\bea
\Delta m_s^{\rm SM}({\rm UTfit}) &=& 21.5 \pm 2.6 ~{\rm ps}^{-1}, \nl
\Delta m_s^{\rm SM}({\rm CKMfit}) &=& 21.7^{+5.9}_{-4.2} ~{\rm ps}^{-1},
\label{dms:SM}
\eea
which are obtained from global fits,
although the experimental measurements in (\ref{dms:exp}) are slightly lower.
Therefore (\ref{dms:exp}) impose strong constraints which predict large
$b-s$ mixings~\cite{Baek:BsBs}.

Another $b \to s$ dominating processes, $B \to \pi K$ decays, have been extensively
studied~\cite{BFRS}.
The current measurements of branching ratios (BRs) and CP-asymmetries (CPAs)
in the four $B \to \pi K$ channels show some interesting discrepancy from the
SM predictions~\cite{BFRS,Baek:2007yy}.
We argue that this ``$B \to \pi K$ puzzle'' manifests itself in the CP-violating
observables  like the difference between $A_{\rm CP}(B^+ \to \pi^+ K^0)$
and $A_{\rm CP}(B^0 \to \pi^- K^+)$ or $S_{CP}(B^0 \to \pi^0 K^0)$ from its SM predictions.
The puzzle can be solved if we introduce NP in the electroweak penguin
sector~\cite{GNK}.

We'd like to stress that even with the constraint given by (\ref{dms:exp})
there are still much room for NP contributions in $b \to s$ transitions. We
demonstrate that the CPAs in $B \to KK$ decays can be very different from
the values expected in the SM~\cite{BLMV}. In addition, if the NP appears
in the electroweak penguin sector as required by the $B \to \pi K$ puzzle,~\cite{GNK}
the predictions in the two modes, $B \to K^+ K^-$ and $B \to K^0 \bar{K}^0$, can
be very different.

\section{$B_s-\bar{B}_s$ mixing}
\label{sec:BsBs}

We consider the implications of (\ref{dms:exp}) on an MSSM
scenario with large mixing in the LL and/or RR sector. We do not consider
flavor mixing in the LR(RL) sector because they are i) are already strongly
constrained by $BR(B \to X_s  \gamma)$ and ii) therefore relatively insensitive to
$B_s - \ol{B}_s$ mixing. We neglect mixing between the 1st and
2nd generations which are tightly constrained by $K$ meson decays and
$K - \ol{K}$ mixing, and mixing between the 1st and 3rd generations
which is also known to be small by the measurement of $B_d - \ol{B}_d$ mixing.
And the down-type squark mass matrix is given by
\bea
 M^2_{\wt{d},LL}
=\l(\ba{ccc}
 \wt{m}^{d,2}_{L_{11}} & 0 & 0 \\
 0 & \wt{m}^{d,2}_{L_{22}} & \wt{m}^{d,2}_{L_{23}} \\
 0 & \wt{m}^{d,2}_{L_{32}} & \wt{m}^{d,2}_{L_{33}} \\
\ea
\r), \quad
 M^2_{\wt{d},LR(RL)} \equiv 0_{3\times 3}.
\eea
The  $M^2_{\wt{d},RR}$ can be obtained from $M^2_{\wt{d},LL}$
by exchanging $L \leftrightarrow R$.
We note that this kind of scenario is orthogonal to the one with flavor
violation controlled only by CKM matrix
minimal flavor violation model~\cite{MFV},
where large flavor violation in $s-b$ is impossible a priori.
Defining
\bea
  M_{12}^s \equiv M_{12}^{s,{\rm SM}} (1+R),
\eea
we obtain the following constraint,
\bea
 |1+R| = 0.77 ^{+0.02}_{-0.01}({\rm exp}) \pm 0.19 ({\rm th}).
\label{eq:R}
\eea

\begin{figure}[tbh]
\begin{center}
\psfrag{mmssLL}{$m_{ \wt{s}_L} ({\rm TeV})$}
\psfrag{ttLL}{$\te_L$}
\subfigure[]{
\includegraphics[width=0.2\textwidth]{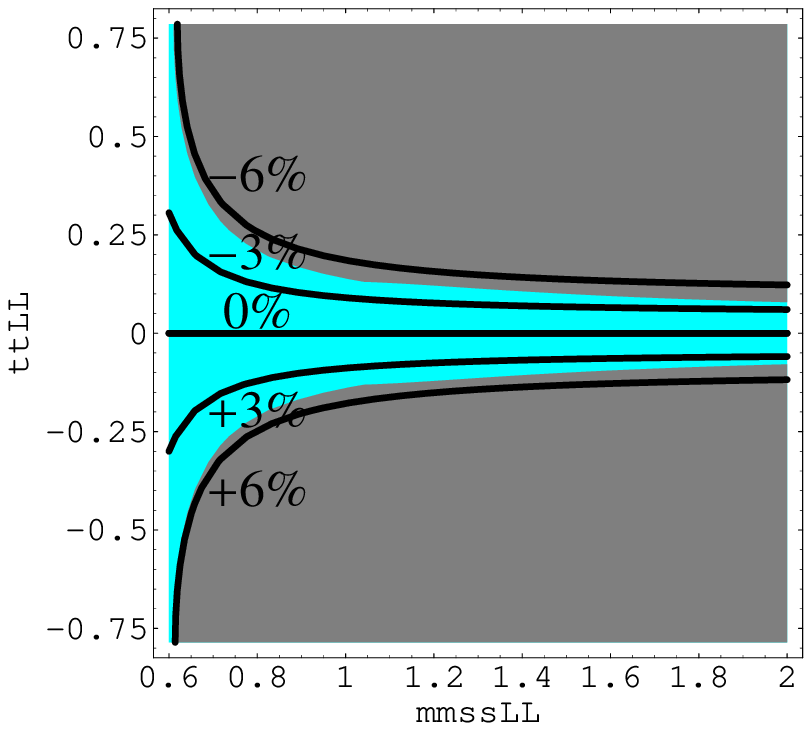}
\label{fig:teL-msL-a}
}
\subfigure[]{
\includegraphics[width=0.2\textwidth]{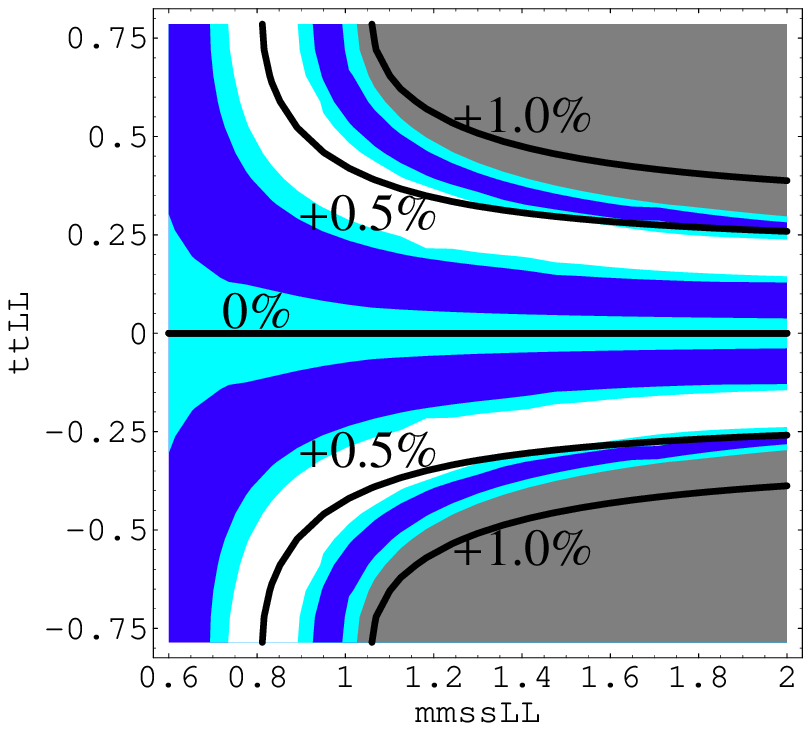}
\label{fig:teL-msL-b}
}
\end{center}
\caption{
Contour plots for $|1+R|$ in ($m_{\wt{s}_L}$,$\te_L$) plane.
Sky blue region represents
2$\sigma$ allowed region ($0.39 \le |1+R| \le 1.15$),
blue 1$\sigma$ allowed region ($0.58 \le |1+R| \le 0.96$),
and white (grey) region is excluded at 95\% CL by giving too small (large)
$\Delta m_s$.
The labeled thick lines represent
the constant
$\Big(BR^{\rm tot}(B \to X_s  \gamma)-BR^{\rm SM}(B \to X_s  \gamma)\Big)/
BR^{\rm SM}(B \to X_s  \gamma)$ contours.
Only LL mixing
is assumed to exist. The fixed parameters
are $m_{\wt{g}}=0.5$ (TeV),  $m_{\wt{b}_L}=0.5$ (TeV),
(a) $\de_L$=0, (b) $\de_L=\pi/2$.
}
\label{fig:teL-msL}
\end{figure}

The larger the mass splitting between
$\wt{s}$ and $\wt{b}$, the larger the SUSY contributions are.
Therefore we expect
that $\Delta m_s^{\rm exp}$ constrains the mass splitting when the mixing
angle $\te_{L(R)}$
is large. This can be seen in Figure~\ref{fig:teL-msL} where we show
filled contour plots for $|1+R|$ in ($m_{\wt{s}_L}$,$\te_L$) plane:
sky blue region represents
2$\sigma$ allowed region ($0.39 \le |1+R| \le 1.15$),
blue 1$\sigma$ allowed region ($0.58 \le |1+R| \le 0.96$),
and white (grey) region is excluded at 95\% CL by giving too small (large)
$\Delta m_s$.
For these plots we assumed that
only LL mixing exists and fixed $m_{\wt{g}}=0.5$ TeV, $m_{\wt{b}_L} = 0.5$ TeV.
In Figure~\ref{fig:teL-msL-a}, we fixed $\de_L = 0$. We can see that
the SUSY interferes with the SM contribution constructively
({\it i.e.} the SUSY contribution has the same sign with the SM),
and when the mixing angle
is maximal, {\it i.e.} $\te_L = \pm\pi/4$, $m_{\wt{s}_L} - m_{\wt{b}_L}$
cannot be
greater than about 150 GeV. In Figure~\ref{fig:teL-msL-b},
we set $\de_L = \pi/2$.
The SUSY contribution can interfere destructively
({\it i.e.} in opposite sign)
with the SM and much larger mass splitting
is allowed. Therefore we can see that the allowed parameters
are sensitive to the
CPV phase.
We see that the $B(b \to s \gamma)$ constraint is not important in this case.

\begin{figure}[tbh]
\begin{center}
\psfrag{mmssLL}{$m_{ \wt{s}_L} ({\rm TeV})$}
\psfrag{ddLL}{$\de_L$}
\subfigure[]{
\includegraphics[width=0.2\textwidth]{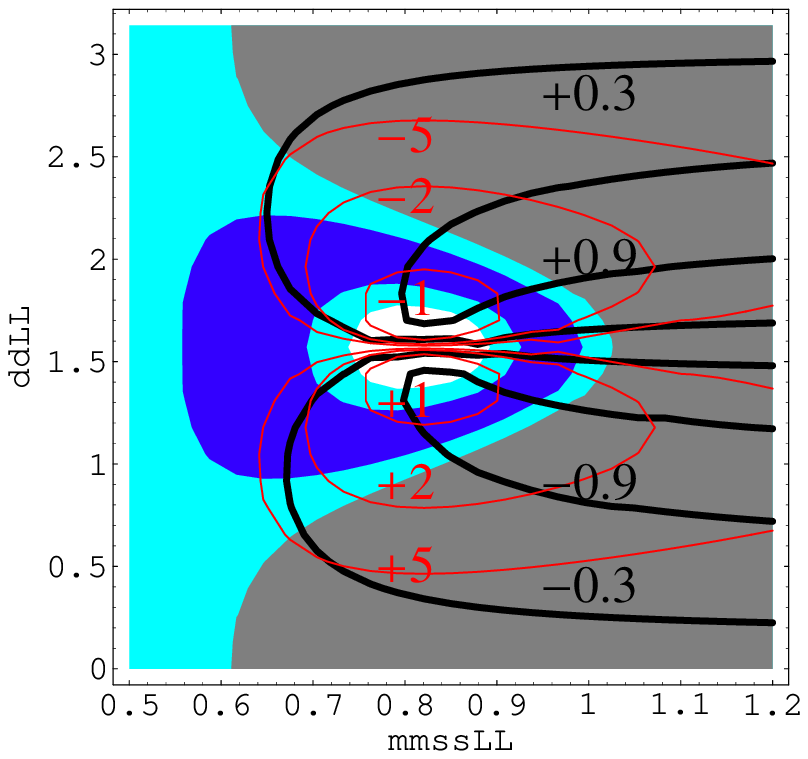}
\label{fig:msL-deL-a}
}
\subfigure[]{
\includegraphics[width=0.2\textwidth]{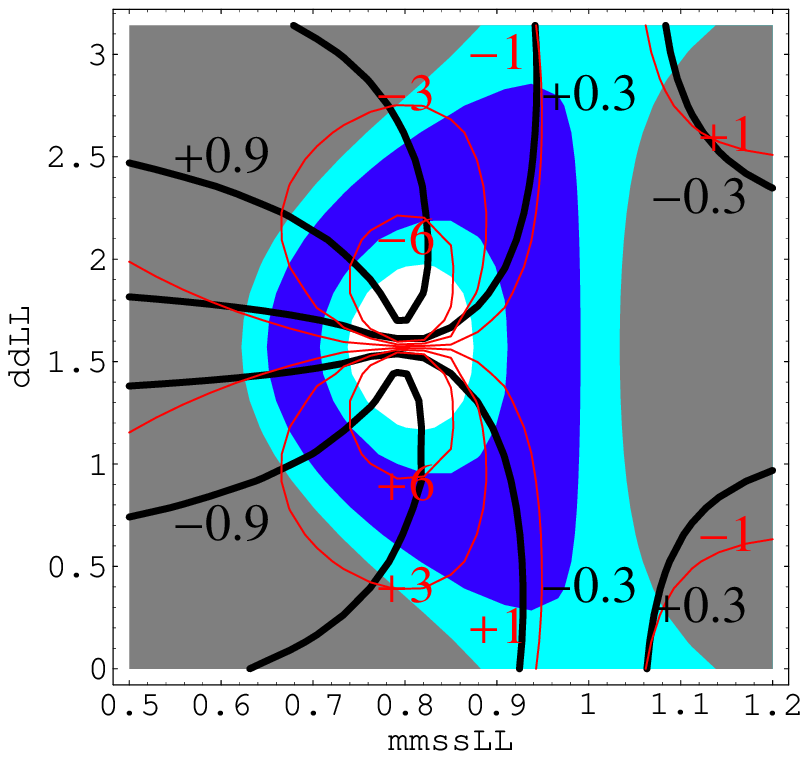}
\label{fig:msL-deL-b}
}
\end{center}
\caption{Contour plots for $|1+R|$ in ($m_{\wt{s}_L}$,$\de_L$) plane.
The $S_{\psi \phi}$ predictions are also shown as thick contour lines.
The thin red lines are constant $A_{SL}^s[10^{-3}]$ contours assuming
${\rm Re}(\Gamma_{12}^s / M^s_{12})^{\rm SM} = -0.0040$.
(a) Only LL mixing is assumed to exist.
We fixed $m_{\wt{g}}=m_{\wt{b}_L}=0.5$ TeV, $\de_L = \pi/4$.
(b) Both LL and RR mixing are assumed to exist simultaneously.
We fixed $m_{\wt{g}}=2$ TeV,
$m_{\wt{b}_L}=m_{\wt{b}_R} = 1$ TeV,
$m_{\wt{s}_R} = 1.1$ TeV,
$\te_R = \pi/4$,
$\de_L = \pi/4$, and
$\de_R = \pi/2$.
The rest is the same with Figure~\ref{fig:teL-msL}.
}
\label{fig:msL-deL}
\end{figure}

The CPV phase in the  $B_s - \ol{B}_s$ mixing amplitude will be measured
at the LHC in the near future through the time-dependent CP
asymmetry
\bea
 \frac{\Gamma(\ol{B}_s(t) \to \psi\phi)-\Gamma(B_s(t) \to \psi\phi)}
{\Gamma(\ol{B}_s(t) \to \psi\phi)+\Gamma(B_s(t) \to \psi\phi)}
\equiv S_{\psi\phi} \sin(\Delta m_s t).
\eea
In the SM, $S_{\psi\phi}$ is predicted to be very small,
$S_{\psi\phi}^{\rm SM} = -\sin 2 \beta_s = 0.038 \pm 0.003$
($\beta_s \equiv \arg[(V_{ts}^* V_{tb})/(V_{cs}^* V_{cb})]$).
If the NP has additional CPV phases, however, the prediction
\bea
 S_{\psi\phi} = -\sin(2 \beta_s + \arg(1 + R))
\eea
can be significantly different from the SM prediction.

In Figure~\ref{fig:msL-deL}, we show $|1+R|$ constraint and the prediction
of $S_{\psi\phi}$  in $(m_{\wt{s}_L},\de_L)$ plane.
However, the $B\to X_s \gamma$ prediction is not shown from now on
because it is irrelevant as mentioned above.
For Figure~\ref{fig:msL-deL-a}, we assumed the scenario with LL mixing only
and maximal mixing $\te_L = \pi/4$.
We fixed $m_{\wt{g}}=0.5$ TeV, $m_{\wt{b}_L} = 0.5$ TeV.
For Figure~\ref{fig:msL-deL-b}, we allowed both LL and RR mixing simultaneously,
while fixing
$m_{\wt{g}}=2$ TeV,
$m_{\wt{b}_L}=m_{\wt{b}_R} = 1$ TeV,
$m_{\wt{s}_R} = 1.1$ TeV,
$\te_R = \pi/4$,
$\de_L = \pi/4$, and
$\de_R = \pi/2$.
In both cases we can see that large $S_{\psi\phi}$ is
allowed for large mass splitting
between $m_{\wt{b}_L}$ and $m_{\wt{s}_L}$.
At the moment, $S_{\psi\phi}$
can take any value in the range $[-1,1]$ even after
imposing the current $\Delta m_s^{\rm exp}$ constraint.

\section{The $B \to \pi K$ puzzle}
The $B \to \pi K$ decays, dominated by $b \to s$ transitions, are
one of the most promising candidates where large NP contributions can
be probed. The current data shown in Tab.~\ref{tab:data}, can be analyzed
using the diagrammatic amplitudes~\cite{Baek:2007yy}:
\bea
\label{fulldiagrams}
A^{+0} &\!\!=\!\!& -P'_{tc} + P'_{uc} e^{i\gamma} -\frac13
\pewcp ~, \nl
\sqrt{2} A^{0+} &\!\!=\!\!& -T' e^{i\gamma} -C' e^{i\gamma}
+P'_{tc} \nl
& & ~~~~ -~P'_{uc} e^{i\gamma} -~\pewp -\frac23 \pewcp ~, \nl
A^{-+} &\!\!=\!\!& -T' e^{i\gamma} + P'_{tc} -P'_{uc}
e^{i\gamma} -\frac23 \pewcp ~, \nl
\sqrt{2} A^{00} &\!\!=\!\!& -C' e^{i\gamma} - P'_{tc} +P'_{uc}
e^{i\gamma} - \pewp -\frac13 \pewcp .
\eea

\begin{table}[tbh]
\center
\begin{tabular}{cccc}
\hline
\hline
Mode & $BR[10^{-6}]$ & $A_{\sss CP}$ & $S_{\sss CP}$ \\ \hline
$B^+ \to \pi^+ K^0$ & $23.1 \pm 1.0$ & $0.009 \pm 0.025$ & \\
$B^+ \to \pi^0 K^+$ & $12.8 \pm 0.6$ & $0.047 \pm 0.026$ & \\
$\bd \to \pi^- K^+$ & $19.7 \pm 0.6$ & $-0.093 \pm 0.015$ & \\
$\bd \to \pi^0 K^0$ & $10.0 \pm 0.6$ & $-0.12 \pm 0.11$ &
$0.33 \pm 0.21$ \\
\hline
\hline
\end{tabular}
\caption{Branching ratios, direct CP asymmetries $A_{\sss CP}$, and
mixing-induced CP asymmetry $S_{\sss CP}$ (if applicable) for the four
$\btopik$ decay modes. }
\label{tab:data}
\end{table}

Neglecting $P'_{uc}$ and $\pewcp$ which are expected to give
small contributions, we can fit (\ref{fulldiagrams}) to the data
in Tab.~\ref{tab:data}.
The ratio $|C'/T'| = 1.6 \pm 0.3$ is required here (we stress
that correlations have been taken into account in obtaining this
ratio). This value is much larger than the naive estimates, 
the NLO pQCD prediction \cite{PQCD_NLO},
$|C'/T'| \sim 0.3$, and the maximal SCET (QCDf) prediction
\cite{SCET}, $|C'/T'| \sim 0.6$. Thus, if one takes this
theoretical input seriously -- as we do here -- this shows explicitly
that the $\btopik$ puzzle is still present, at $\sim 3\sigma$
level. (The abnormally large value of $|C'/T'| = 1.6 \pm 0.3$ found
here is partially due to $S_{\sss CP}$.  Without it we obtain $|C'/T'|
= 0.8 \pm 0.1$.) In Ref.~\cite{Baek:2004} (2004), $|C'/T'| = 1.8 \pm
1.0$ was found. We thus see that the puzzle has gotten much worse in
2006. In passing we note that the similar problem in $B \to \pi \pi$ decays
can be solved if we can separate the $P_{uc}$ component from the
$T$ and $C$ amplitude using, for example, the measurements of
 $B \to KK$ decays~\cite{Baek:2007}.


If we include NP,
the NP contribution in the electroweak penguin amplitude,
$\ApNPcomb$, fits the data best: $\chi^2_{min}/d.o.f. = 0.6/3~(90\%)$.
For this fit, we set other NP amplitudes to be zero.
This is the same conclusion as that found in Ref.~\cite{Baek:2004}.
Thus, not only is the $\btopik$ puzzle still present, but it is still
pointing towatds the same type of NP, $\ApNPcomb \ne 0$. For this (good) fit, we find
$|T'/P'| = 0.09$, $|{\cal A}^{\prime, comb}/P'| = 0.24$, $\Phi' =
85^\circ$. We therefore find that the NP amplitude must be sizeable,
with a large weak phase.

\section{Large SUSY contributions to $B \to KK$ decays}
In the SM the $B_s \to KK$ decays can be parameterized as
\bea {\cal A}(\bs \to K^+ K^-) & \simeq & V_{ub}^* V_{us}[T'+
  (P'_u - P'_t)] \nl
  &&+ V_{cb}^* V_{cs} (P'_c - P'_t) \nl
& \equiv & V_{ub}^* V_{us} T^{s\pm} + V_{cb}^* V_{cs} P^{s\pm} ~, \nl
{\cal A}(\bskkneut) &\simeq& V_{ub}^* V_{us} T^{s0} + V_{cb}^* V_{cs} P^{s0}
~.
\eea
The amplitudes $P^{s\pm}$, $P^{s0}$, $T^{s\pm}$ and $T^{s0}$ can
be determined from the measurements of $B_d^0 \to K^0 \bar K^0$ decay~\cite{DMV}.
The amplitude for $B_d^0 \to K^0 \bar K^0$ can be written
\bea
{\cal A}(B_d^0 \to K^0 \bar K^0) \simeq V_{ub}^* V_{ud} T^{d0} + V_{cb}^* V_{cd} P^{d0}
~.
\eea
The three unknown physical quantities in $T^{d0}$ and $P^{d0}$
are determined from the three conditions: 
{\it i)}
\bea 
BR(B_d \to K^0 \bar K^0) = (0.96\pm 0.25) \times 10^{-6},
\eea
{\it ii)} a quantity free from IR cutoff in QCDF~\cite{QCDf1,QCDf2},
\beq
\Delta_d = (1.09\pm 0.43) \times 10^{-7} + i (-3.02 \pm 0.97) \times
10^{-7} ~{\rm GeV},
\label{Deltad}
\eeq
where $\Delta_d \equiv T^{d0} - P^{d0}$, and 
{\it iii)} the fact that
only values $-0.2
\le A_{dir}^{d0} \le 0.2$ are consistent with the measured value of
$BR(B_d \to K^0 \bar K^0)$ and the theoretical value of $\Delta_d$~\cite{DMV}.
From these conditions we obtain
\bea
|T^{d0}| &=& (1.1 \pm 0.8) \times 10^{-6}~{\rm GeV}, \nl
|P^{d0}/T^{d0}| &=& 1.2 \pm 0.2 ~, \nl
{\rm arg}(P^{d0}/T^{d0}) & = & (-1.6 \pm 6.5)^{\circ} ~.
\eea

\begin{figure}
\begin{center}
\psfrag{Adirs0}{\tiny \hspace{-1cm} $\stackrel{A_{\rm dir}(\bskkneut)\times 10^{2}}{}$} 
\psfrag{Amixs0}{\tiny \hspace{-1cm} $\stackrel{A_{\rm mix}(\bskkneut)\times 10^{2}}{}$}
\psfrag{Adirspm}{\tiny \hspace{-0.8cm} $\stackrel{A_{\rm dir}(B_s^0 \to K^+ K^-)}{}$} 
\psfrag{Amixspm}{\tiny \hspace{-0.8cm} $\stackrel{A_{\rm mix}(B_s^0 \to K^+ K^-)}{}$} 
\psfrag{Adird0}{}
\includegraphics[width=8cm]{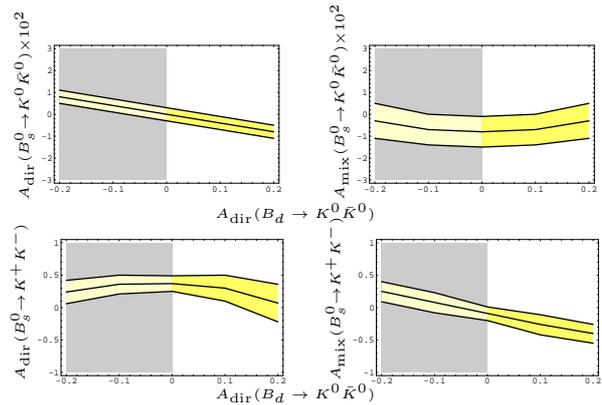}
\Text(-150,2)[lb]{\tiny $A_{\rm dir}(B_d \to K^0 \bar K^0)$}
\Text(-150,70)[lb]{\tiny $A_{\rm dir}(B_d \to K^0 \bar K^0)$}
%
\end{center}
\caption{SM predictions for the CP asymmetries in $\bskkneut$ (up) and
$B_s^0 \to K^+ K^-$ (down) as a function of $A_{\rm dir}(B_d \to K^0 \bar K^0)$. 
As explained in
the text, the preferred range is the non-shadowed half of the plots
[$A_{\rm dir}(B_d \to K^0 \bar K^0)\geq 0$].}
\label{plotACPs}
\end{figure}

Now we can relate the parameters in $B_d \to K^0 \bar K^0$
decays to those in the $B_s \to KK$ decays 
using SU(3) symmetry. We impose the factorizable SU(3)-breaking
effect
\beq f = {M_{\bs}^2 F_0^{\bs \to \kbar}(0) \over M_{\bd}^2
F_0^{\bd \to \kbar}(0)} = 0.94 \pm 0.20 ~. \eeq
We can predict the observables in $B_s \to KK$ decays shown
in Fig.~\ref{plotACPs}

The NP contribution can be parameterized as
\bea
 {\cal A}(B_s^0 \to K^+ K^-) &=& {\cal A}^{s\pm}_{\sss SM} + {\cal A}^u e^{i
\Phi_u}, \nl
{\cal A}(\bskkneut) &=& {\cal A}^{s0}_{\sss SM} + {\cal A}^d e^{i \Phi_d} ~.
\label{AmpBsKK}
\eea
If the NP conserves isospin, we have ${\cal A}^u = {\cal A}^d$ and $\Phi_u =
\Phi_d$, but in general this need not be the case. Especially in our NP
model~\cite{BLMV} described in Section~\ref{sec:BsBs}, there can be large
isospin violation~\cite{GNK}. To see how large the NP contributions can be
we scanned in the following SUSY parameter space:
\begin{itemize}

\item
$m_{\tilde{u}_L}=m_{\tilde{d}_{L,R}}=m_{\tilde{b}_{L,R}}=m_{\tilde{g}}=
250 ~{\rm GeV}$

\item
$250 ~{\rm GeV} < m_{\tilde{u}_R}, m_{\tilde{s}_{R,L}} < 1000 ~{\rm
GeV}$

\item $-\pi < \delta_{L,R} < \pi$

\item $-\pi/4 < \theta_{L,R} < \pi/4$

\end{itemize}
We imposed
$BR(B\to X_s\gamma)=(3.55\pm 0.26)\times 10^{-4}$
and 
$\Delta m_s$ constraints considered in Section~\ref{sec:BsBs}.

\begin{figure}
\begin{center}
\psfrag{BR}{\hspace{-1cm} \tiny $\stackrel{BR(K^+K^-)\times 10^6}{}$}
\psfrag{Amix}{\hspace{-1cm} \tiny $\stackrel{A_{\rm mix}(K^+K^-)}{}$}
\psfrag{Adir}{}
\includegraphics[width=8cm]{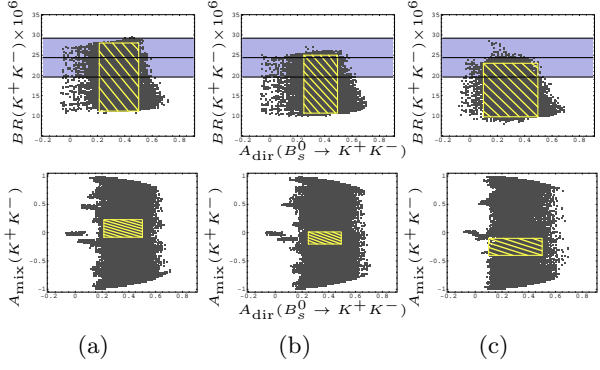}
\Text(-140,0)[lb]{\tiny $A_{\rm dir}(\bskk)$}
\Text(-140,60)[lb]{\tiny $A_{\rm dir}(\bskk)$}
%
\Text(-200,-15)[lb]{\small (a)}
\Text(-125,-15)[lb]{\small (b)}
\Text(-50,-15)[lb]{\small (c)}
\end{center}
\caption{\small Predictions, in the form of scatter plots, for the
correlations between $BR(\bskk)-A_{\rm dir}(\bskk)$ (up) and
$A_{\rm mix}(\bskk)-A_{\rm dir}(\bskk)$ (down) in the presence of
SUSY, for a) $A_{dir}^{d0}=-0.1$, (b) $A_{dir}^{d0}=0$ and (c)
$A_{dir}^{d0}=0.1$. The dashed rectangles correspond to the SM
predictions. The horizontal band shows the experimental value for
$BR(\bskk)$ at  $1\sigma$.}
\label{plotBK+K-}
\end{figure}

\begin{figure}
\begin{center}
\psfrag{BR}{\hspace{-1cm} \tiny $\stackrel{BR(K^0 \bar K^0)\times 10^6}{}$}
\psfrag{Amix}{\hspace{-1cm} \tiny $\stackrel{A_{\rm mix}(K^0 \bar K^0)}{}$}
\psfrag{Adir}{}
\includegraphics[width=8cm]{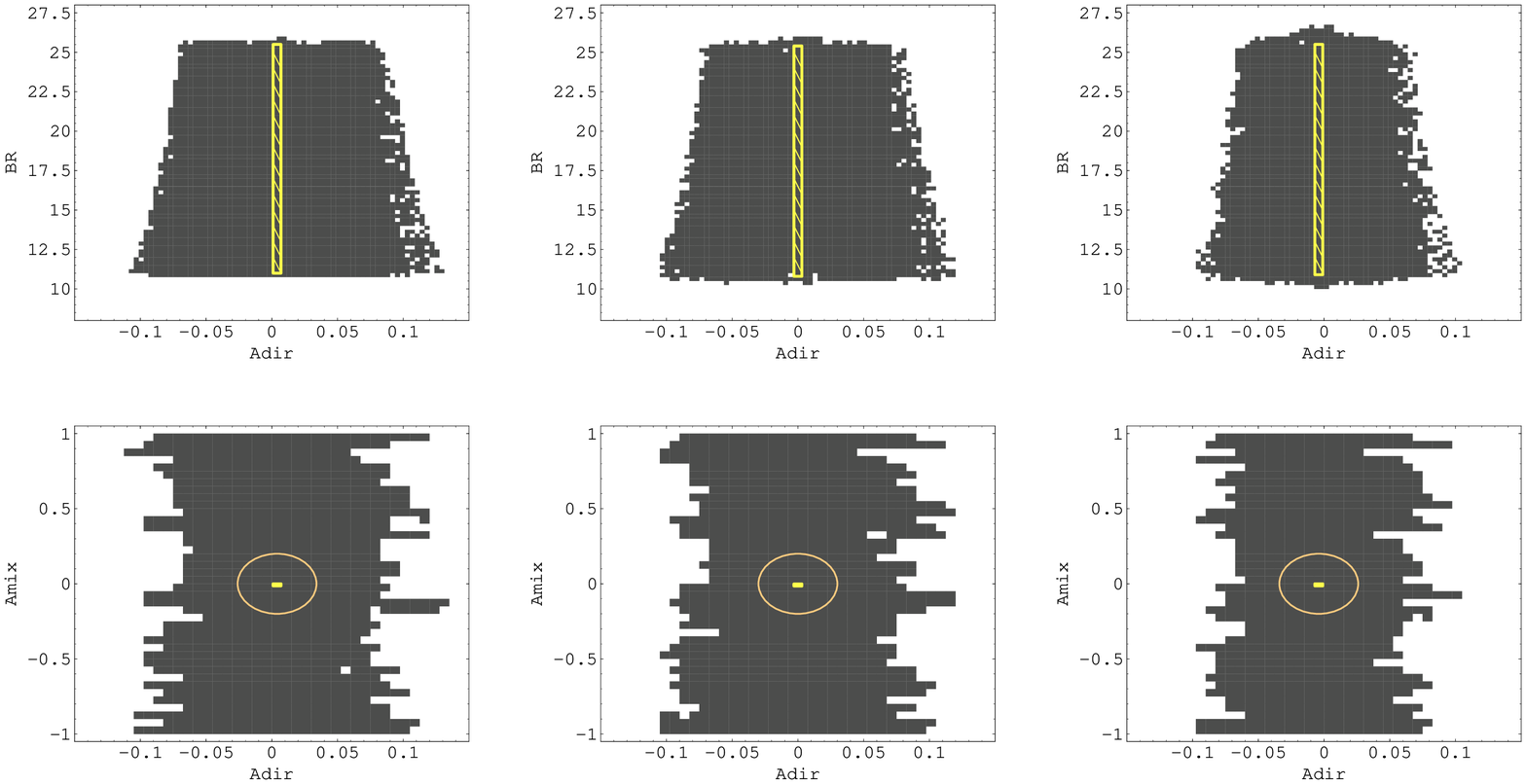}
\Text(-140,0)[lb]{\tiny $A_{\rm dir}(\bskkneut)$}
\Text(-140,60)[lb]{\tiny $A_{\rm dir}(\bskkneut)$}
%
\Text(-200,-15)[lb]{\small (a)}
\Text(-125,-15)[lb]{\small (b)}
\Text(-50,-15)[lb]{\small (c)}
\end{center}
\caption{\small Predictions, in the form of scatter plots, for the
correlations between $BR(\bskkneut)-A_{\rm dir}(\bskkneut)$ (up) and
$A_{\rm mix}(\bskkneut)-A_{\rm dir}(\bskkneut)$ (down) in the
presence of SUSY, for (a) $A_{dir}^{d0}=-0.1$, (b) $A_{dir}^{d0}=0$
and (c) $A_{dir}^{d0}=0.1$. The dashed rectangles correspond to the
SM predictions. These are quite small in the three lower plots, so
they are indicated by a circle.} 
\label{plotBK0K0}
\end{figure}

As can be seen in Figs.~\ref{plotBK+K-} and \ref{plotBK0K0}, there
can be huge deviations from the SM predictions, in 
$A_{\rm mix}(B_s^0 \to K^+ K^-)$,
$A_{\rm dir,mix}(B_s^0 \to K^0 \bar K^0)$.

\section{Conclusions}
We have seen that the $\Delta m_s^{\rm exp}$ gives strong
constraints on large $b-s$ mixing in NP. On the other hand
the nonleptonic $B \to \pi K$ decays seem to require NP contributions.
We have shown that even with the $\Delta m_s^{\rm exp}$
constraint the NP still allows large $b-s$ leaving observable
effects, for example, in $B_s \to K K$ decays.

\vspace{1cm}
{\bf Acknowledgements}: \\
The author thanks David London, Joaquim Matias, and Javier Virto
for collaborations. This work
was supported by the Korea Research Foundation Grant
funded by the Korean Government (MOEHRD) 
No. KRF-2007-359-C00009.

%
%

\end{document}